\documentclass[runningheads]{llncs}
\usepackage[T1]{fontenc}
\usepackage{lineno}
\newcommand{\reg}[1]{\textrm{#1} }

\newcommand{\rep}{\textrm{rep}}

\newcommand{\concat}{\ensuremath{\odot}}
\newcommand{\union}{\ensuremath{\oplus}}

\newcommand{\tabD}{\ensuremath{\textit{tabD}\,}}

\begin{document}

\title{A Program That Simplifies Regular Expressions}
\subtitle{(Tool paper)}

\author{Baudouin {Le Charlier} \orcidID{0000-0002-2222-4203} }

\authorrunning{B. {Le Charlier}}

\institute{UCLouvain--ICTEAM\\
\email{baudouin.lecharlier@uclouvain.be}}

\maketitle             
\begin{abstract}

This paper presents the main features of a system that aims to transform regular expressions into shorter equivalent expressions. The system is also capable of computing other operations useful for simplification, such as checking the inclusion of regular languages. The main novelty of this work is that it combines known but distinct ways of representing regular languages into a global unified data structure that makes the operations more efficient. In addition, representations of regular languages are dynamically reduced as operations are performed on them. Expressions are normalized and represented by a unique identifier (an integer). Expressions found to be equivalent (i.e. denoting the same regular language) are grouped into equivalence classes from which a shortest representative is chosen.

The article briefly describes the main algorithms working on the global data structure. Some of them are direct adaptations of well-known algorithms, but most of them incorporate new ideas, which are really necessary to make the system efficient. 
Finally, to show its usefulness, the system is applied to some examples from the literature. Statistics on randomly generated sets of expressions are also provided.

\end{abstract}

\section{The Background: a Global Data Structure}
\label{background}

Regular expressions, regular languages, and deterministic finite automata are well-known (see, e.g. \cite{Parsing,Brzozowski,Conway,Kozen94}). In this paper, symbols in chains are lower case letters. The symbols $0$  and $1$ denote the empty set, and the set only containing the empty chain, respectively.
The system works on \emph{normalized expressions}. Letters and symbols $0$ and $1$ are normalized expressions. A normalized iteration is of the form \reg{$E$*,} where $E$ is different from $0$ and $1$ and is not an iteration; a normalized concatenation is of the form \reg{$E_1\,.\, E_2$} (also written \reg{$E_1\,E_2$),}  where $E_1$ and $E_2$ are different from $0$ and $1$, and $E_1$ is not a concatenation; a normalized union is of the form $E_1 + \ldots + E_n$, where $n \ge 2$, and all expressions $E_i$ are different from $0$ and are not unions; moreover the sequence $E_1, \ldots, E_n$ is strictly sorted. Thus, we assume a total order on normalized expressions. For efficiency reasons, this order is implementation dependent. Arbitrary regular expressions can be mapped in a unique way to a normalized expression, thanks to three operations $E_1\union E_2$, $E_1\concat E_2$, and $E^{\displaystyle\star}$, which compute normalized expressions equivalent to $E1 + E2$, $E_1 . E_2$, and \reg{$E$*,} assuming  that $E_1$, $E_2$, and $E$ are normalized. It is easy to see that any two arbitrary regular expressions that can be shown equivalent using the Kleene's classical axioms\footnote{see \cite{Conway}, page 25} for $0$, $1$, $.$, and $+$, except distributivity, as well as the axiom $\reg{($E$*)*}=\reg{$E$*}$, and identities $\reg{$0$*} = 1$ and $\reg{$1$*} = 1$, are mapped onto the same normalized expression.

The system uses a global data structure, called \emph{the background}, that contains a set of normalized expressions and a set of equations relating (some of) the regular expressions to their derivatives (see \cite{Brzozowski,Conway}). Following  \cite{Conway}, such an equation can be written $E\:=\:o_E + \ldots + x.E_x + \ldots$, where $o_E\in\{0, 1\}$, $x$ is a letter, and every $E_x$ is a normalized expression present in the background. The background evolves over time. Actually, all operations executed by the system use expressions and/or equations present in the background to create new expressions and equations, which are added to it. The background also maintains an equivalence relation between the normalized expressions it contains. As expected, expressions in the same equivalence class must denote the same regular language. Moreover, in each class, a shortest expression is chosen as \emph{the representative} of the class. We note $\rep(E)$ the representative of the equivalence class of $E$. In an equation as above, it is required that $E = \rep(E)$ and $E_x = \rep(E_x)$ for every letter $x$.  It is also natural to require that no two equations may use the same left part $E$ or the same right part $o_E + \ldots + x.E_x + \ldots$. Let us call this the \textit{invariant} of the background. But, as shown in Section \ref{simplification}, this condition may be violated when two expressions are found equivalent. Thus, there is another global operation, called \textit{normalize}, which enforces the condition, when needed. This means that some equivalence classes of expressions are merged, choosing a shortest representative, and that some sets of equations are replaced by a single  new one, using new representatives. Replaced equations are discarded from the background.\vspace{-1em}

\section{Implementation of the Background and its Operations}
\label{implementation}

\vspace{-1em}The implementation of the background almost only uses integers, arrays of integers and arrays of arrays of integers, possibly ``encapsulated'' into objects, for readability.\footnote{The program is written in Java, and can be compiled with any version of it, including Java 1.0. It could be readily re-coded in most imperative programming language.} Normalized regular expressions are (uniquely) identified by an integer (int). There is a unique array (of arrays of integers) giving access to all expressions in the background. The identifier of an expression $E$ gives access through this array to an array of integers containing the identifiers of the direct sub-expressions of $E$. The length of the main array determines the maximum number of expressions present in the background. 

Whenever all identifiers are used for expressions, it is often possible to get rid of some of them, no longer strictly needed for the current task. This is handled by a specialized garbage collector. To make it possible, all identifiers are  distributed into two lists: the used ones and the free ones. These lists, and other needed ones, are made of arrays of integers, allowing us to perform operations such as choosing and removing an identifier, or checking its presence, all in constant time. When operations such as $E_1\union E_2$, $E_1\concat E_2$, and $E^{\displaystyle\star}$ are performed, they receive, as actual arguments, identifiers of expressions. A hashtable is used to check whether the result already exists, or to create it with a free identifier.
When the result is a union, its direct sub-expressions are merged in ascending order on the values of their identifiers, which makes the time complexity of the operation proportional to the number of these sub-expressions. This cannot be obtained by  a priori defining a total ordering on the expressions. In fact, the ordering is defined by the program execution history. 

Expressions in the background are gathered into equivalence classes. To implement this, we use an array of identifiers organized as a Union-Find data structure \cite{GallerF64}. This provides a fast access to the identifier of $\rep(E)$ from the identifier of $E$.

To implement the operation \textit{normalize} of the background, I first explain how equations $E\,=\,o_E + \ldots + x.E_x + \ldots$ are represented. The right part $o_E + \ldots + x.E_x + \ldots$ is represented by an array of identifiers $\tabD$ where $\tabD[0]$ is the identifier of $o_E$ ($0$ or $1$).
The length of $\tabD$ is the number of different letters used by all expressions in the background plus one and $\tabD[i_x]$ is the identifier of $E_x$ (where $i_x$ is 
the rank of $x$ in the set of used letters (starting at $1$)).\footnote{When expressions are normalized, their letters are renamed to use the first ones of the alphabet.} Arrays of identifiers themselves are given an identifier using another hash-table. Additionally, an equation is represented by a pair consisting of the identifiers of its left and right parts. Finally, an identifier is given to each of these pairs, thanks to a third hash table.
This identifier is used to access two arrays where the identifiers of the left and right parts of the equation are put. Let us assume that two expressions $E_1$ and $E_2$ such that $\rep(E_1) \neq \rep(E_2)$ are found equivalent and that $\rep(E_1)$ is shorter than $\rep(E_2)$. Their equivalence classes are merged in the Union-Find structure. Moreover, to maintain the invariant of the background, we replace  $\rep(E_2)$ by $\rep(E_1)$ in every equation where $\rep(E_2)$ occurs. To make it efficiently, the background contains, for every position $i_x$ and every representative $E$ of an equivalence class, a list of all identifiers of arrays of identifiers $\tabD$ such that  $\tabD[i_x]$
 is equal to the identifier of $E$. Assume that the equations contain $n$ occurrences of $\rep(E_2)$. Then the old equations containing  $\rep(E_2)$ can be replaced by new equations using $\rep(E_1)$ instead, in $O(n\times \ell_\tabD)$ where $\ell_\tabD$ is the length of arrays of identifiers  $\tabD$ (on the average, i.e. if the hash-tables work well). ``Renaming'' $\rep(E_2)$ into $\rep(E_1)$, this way, does not maintain the invariant of the background, in general: two equations may have $\rep(E_1)$ as left part, and several equations may have the same right part. Thus, the same process may have to be iterated until the invariant holds anew. This is efficiently done using a stack of pairs of identifiers to be put in the same equivalence class and other lists of identifiers of equations having the same left or right parts. The whole process is guaranteed to terminate since the number of distinct identifiers used by the equations decreases at each iteration.
\section{Simplification and Other Algorithms}
\label{simplification}

The size (or length) of an expression is defined as the number of symbols (occurrences) it is made of, excluding parentheses.
I take the viewpoint that simplifying an expression just means finding another expression that is shorter and denotes the same regular language. The idea is that a shorter expression is easier to understand than a larger one, in general. A more elaborated simplicity measure is proposed in \cite{Stoughton}, aiming notably at limiting the star height (see, e.g. \cite{Lombardy}) of the expressions. With this measure, the expression \reg{1 + a(a + b)*} is considered simpler than \reg{(a b*)*,} a shorter one. It would be possible to use this measure in my system at the price of losing some efficiency.

In theory, simplifying a regular expression can be done, by hand, using Kleene's axioms, possibly extended with Salomaa's rule \cite{Parsing,Conway} or the more logical rules proposed in \cite{Kozen94}. However this requires some ``expertise'' \cite{Conway,Kozen94}. For relatively large expressions, the number of possibilities to try makes the approach impractical. In \cite{Stoughton}, an approach is proposed where a strictly decreasing sequence of expressions is constructed by choosing, at each step, a smaller expression from a large set of expressions equivalent to the preceding one. This approach also is inefficient for large expressions and unable to simplify an expression such as \reg{c* + c*a(b + c*a)*c*} into \reg{(c + ab*)*} because this requires building an intermediate expression strictly greater that the first one (namely, \reg{(1 + c*ab*(c*ab*)*)c*}).

In this work, I suggest using other kinds of algorithms that are ``more deterministic'', making them able to work on larger expressions. They also take advantage of the background, which allows them to reuse the results of previous computations. The general simplification algorithm of an expression works as follows. The expression first is normalized. Then the normalized expression is put on a stack, with its sub-expressions, the shortest ones on the top. Of course, identifiers are used to represent expressions on the stack. (And, in general, ``expression'' means ``identifier of expression'', below.)
Some sub-expressions may have been put on the stack and simplified previously. They are not put on the stack again.
Then sub-expressions are removed from the stack one by one and processed as follows.

Let $E$ be the current (sub-)expression. It is first pre-simplified by replacing its direct sub-expressions by their representative in the background. In many cases this results in a much shorter expression $E'$. Then a complete set of equations is computed for the pre-simplified expression. We say that a set of equations is \emph{complete} if every expression used by its right part is the left part of an equation in the set. A complete set of equations for $E'$ must contain an equation of which the representative of $E'$ is the left part. It is equivalent to a deterministic finite automaton (DFA) for $E'$. The  set of equations is computed using derivatives (see, e.g. \cite{Brzozowski,Conway}). Our algorithm uses a notion of partial derivative similar to \cite{Antimirov} and is efficient thanks to the use of normalized expressions. The main algorithm does not compute the derivatives one by one as in \cite{Conway} but it uses an array of identifiers, as an accumulator to compute the right part of an equation, and another accumulator to progressively compute suffixes of partial derivatives. The set of all syntactic partial derivatives of a normalized expression is finite and syntactic derivatives are unions of partial derivatives. Thus, termination in ensured. In many cases, however, computing all syntactic derivatives is not needed because their representatives in the background already have a complete set of equations. When a new equation is computed, it is added to the background, which is then normalized. Thanks to normalization the size of the set of equations actually computed for $E'$ can be much smaller than the set of equations that would be built by strictly using syntactic derivatives of $E'$. Nevertheless, normalization of the background does not guarantee that two distinct representatives denote different regular languages. This can be ensured using Moore's well-known algorithm \cite{Moore}. The system  uses it in three different ways: minimize the set of equations of $E'$, check the equivalence of two or more expressions, or make sure that all representatives that have a complete set of equations denote different regular languages.

Computing a set of equations for $E'$ is often sufficient to determine shorter equivalent expressions: some syntactic derivatives of  $E'$ may denote the same language and be shorter, and they may have representatives still shorter, detected by normalizing the background and minimizing the set of equations. However, it is also possible and sometimes useful to create new expressions from the set of equations for $E'$. As an example (from \cite{Conway}), consider the case where $E = E' = \reg{(ab*a + ba*b)*(1 + ab* + ba*)}$. This expression only has three syntactical derivatives and thus three equations. The minimization of these equations gives only one equation : $E = 1 + a.E + b.E$. Unless $E$ has a shorter representative somewhere in the background, no improvement is obtained. However, applying Salomaa's rule, (see, e.g. \cite{Conway}), we get that $E$ is equivalent to \reg{(a + b)*.} More generally, the system proposes an algorithm to solve equations, which can be applied to those of $E'$. This algorithm is quite different from the classical algorithm explained in \cite{Parsing} since it attempts to find a short expression for $E'$ and performs a depth-first traversal of the set of equations. It uses a number of heuristics to limit the depth of the traversal. Basically, it treats the expressions in the equations as variables (as in \cite{Parsing}) but, sometimes, it may choose to use their actual values to limit the depth of the search.

Alternatively or complementarily to the algorithm above, the system is able to apply to $E'$ simplification rules similar to those in \cite{Stoughton}. They make a lot of use of an algorithm to decide inclusion of a regular language (denoted by $E_1$) into another (denoted by $E_2$). The best method  (\cite{Conway}) seems to be to compute the derivatives of the (extended) expression $E_1 \setminus E_2$. Inclusion holds only if no such derivative contains $1$. The computation often is fast (when inclusion does not hold.) {(Another (related) method proposed in \cite{Antimirov2} is more difficult to implement and less efficient in general.)} Inclusion is notably used to remove redundant sub-expressions in unions and concatenations, decompose concatenations, and compute coverings of unions. More powerful simplifications can be achieved ``under star'' (in the sub-expression of an iteration). To the contrary of \cite{Stoughton}, the algorithms do not transform unions and concatenations in all possible ways, using associativity and commutativity because it is too costly for large expressions. Heuristics to find ``interesting'' groupings of sub-expressions are tried, instead. 

{\vspace{-3em}\begin{table}[t]
\caption{\vspace{-1em}Simplification of random expressions of size 1000 with two letters}\label{sre}
\begin{center}
\vspace{-1em}\begin{tabular}{|r|r|r|r|r|r|r|r|r|r|r|r|r|r|r|r|r|r|}
\hline
\ \textit{algorithms}\ \ && $\ell_N$\ \ && $n_\textit{min}$\ \ && $\ell_\textit{avg}$\ \ & $\ell_{1/4}$\ \  & $\ell_{1/2}$\ \  
& $\ell_{3/4}$\ \  && $t_\textit{avg}$\ \  & $t_{1/4}$\ \  & $t_{1/2}$ \ \ 
& $t_{3/4}$\ \  && \textit{gc}\ \    \\[0.3em]\hline\hline
n && 715 && 0 && 620 & 613 & 637 & 656 && 0.45 & 0.15 & 0.35 & 0.62 && 7  \\\hline
 && 715 && 24 && 269 & 5 & 59 & 614 && 0.91 & 0.26 & 0.5 & 1.01 && 6  \\\hline
s && 715 && 25 && 69 & 4 & 24 & 91 && 0.18 & 0.06 & 0.1 & 0.13 && 2  \\\hline
rS && 714 && 25 && 52 & 4 & 20 & 52 && 0.16 & 0.04 & 0.07 & 0.15 && 2  \\\hline
rsS && 715 && 25 && 49 & 4 & 19 & 50 && 0.17 & 0.04 & 0.07 & 0.16 && 1  \\\hline
frsS && 715 && 25 && 47 & 4 & 19 & 48 && 0.41 & 0.09 & 0.18 & 0.34 && 8  \\\hline
\end{tabular}
\end{center}
\end{table}}

\section{Examples and Statistics}

\vspace{-1em}Let us see how the system deals with some examples from the literature. In \cite{Wood}, regular expressions are obtained from non deterministic finite automata: the expression \reg{(aa + b)a*c(ba*c)*(ba*d + d) + (aa + b)a*d} ($\ell = 38$) is obtained with some strategy, while a shorter one ($\ell = 18$) is obtained for the same automaton, with a better strategy. My system simplifies the long expression to the shorter\\ \reg{(b + aa)(a + cb)*(1 + c)d} ($\ell = 18$), which is also nicer. As explained in Section~\ref{implementation}, the system also accepts expressions of the form $E_1 \setminus E_2$, which is enough to compute other boolean operations on expressions. In \cite{Conway}, Conway asks to compute \reg{(xy* + yx)* $\cap$ (y*x + xy)*} but he presents a solution for \reg{(xy* + yx)* $\setminus$ (y*x + xy)*,} instead. The system gives \reg{(yx + x(1 + y(y*yx)*))*} ($\ell = 18$) and \reg{(yx + x(1 + y(y*yx)*))*xy(y(1 + x))*y} ($\ell = 31$) as respective solutions. The solution proposed by Conway is \reg{(yx)*xx*y(yy*x + xx*y)*yy*}($\ell = 31$), which is further simplified to \reg{(yx)*xx*y(yx + x*y)*y} ($\ell = 23$) by the system.

Table \ref{sre} provides statistics on the accuracy and efficiency of the algorithms. A set of 100 randomly chosen regular expressions of size 1000 using two letters has been generated fairly, i.e. every possible expression has the same probability to be chosen. All expressions are simplified using different variants of the algorithms and statistics are computed on the sizes of simplified expressions and execution times. 
The first column lists the chosen algorithms. Column $\ell_N$ gives the average length of the normalized expressions. Column $n_\textit{min}$ is the number of expressions simplified to \reg{(a + b)*}. Columns $\ell_\textit{avg}$, $\ell_{1/4}$, $\ell_{1/2}$ 
, $\ell_{3/4}$ provide the average length, first quartile, median, and third quartile of simplified expressions. The next four columns give similar information about the execution times (in seconds, on a MacBook Pro Early 2015). Column \textit{gc} is the number of garbage collector calls for all simplifications. The first line is the case where only derivatives are computed (with normalization of the background). On the second line, the only change is that a minimization of all equations is applied to the previous results (in the end).
The next three lines report on the cases where simplication algorithms (s), minimization plus equation solving (rS), or both (rsS) are applied. At the last line (frsS), a factorization algorithm is also independently applied to every pre-simplified expression. We see that rS is slightly better than s. Combining the two brings a little improvement at a small cost. Adding factorization (f) is  still more precise but more than two times less efficient.

\pagebreak
%
%
%
\nocite{Antimirov3,BLC1}

\bibliographystyle{plain}
\bibliography{Biblio.bib}
\pagebreak
\appendix

\section{Downloading and Testing the System}
In order to test the system described in this paper, you should download the dropbox file at the address
\begin{verbatim}https://www.dropbox.com/sh/j0l4yt59k2w6tpi/AADLG9qFGg_RF2QQ3TkDd_usa?dl=0
\end{verbatim}
and unzip it as a new directory. The directory contains a jar file and a sub-directory with some test data. It also contains the file \texttt{how-to-use.pdf} providing information on how to use the program.


\section{More Examples} 

Let us start with an example showing that computing derivatives is sometimes enough to get substantial simplifications.
Consider the following expression:
\begin{center}
\reg{((a + b)a*)* + (a + b(1 + b)b)aa(1 + a)}
\end{center}
Computing its syntactic derivatives gives eight different equations (put in the background) but normalization of the background reduces them to a single one, of the form $E\:=\:1 + a.E + b.E$. Therefore, the expression simplifies to 
\reg{((a + b)a*)*,} the shortest of its syntactic derivatives. No other algorithm is needed: the program computes it with an empty list of algorithms (\verb%''%). Assume that the background contains \reg{(a + b)*} beforehand (with a corresponding equation). Then the program produces \reg{(a + b)*} instead of \reg{((a + b)a*)*} because minimization of the set of all equations is applied once at the end of the simplification. The result \reg{((a + b)a*)*} is obtained only with the option \verb%n%, which prevents the program from applying minimization in the end. The program also produces \reg{(a + b)*,} if it is not in the background beforehand, and any combination of algorithms is used, except \verb%n%, \verb%nr%, \verb%nf%, \verb%nrf%.

Now, let us consider the problem of checking whether a regular expression $E_1$ denotes a language included in the language denoted by another expression $E_2$. We can ask the program to simplify $E_1 \setminus E_2$. For instance:
\begin{center}
\reg{( (a*b)*aaaaaaa* $\setminus$ (a + b)*a(a + b)(a + b)(a + b)(a + b)(a + b))}
\end{center}
The returned results is $0$, which indicates that the difference is the empty set. But, we also can try the converse:
\begin{center}
\reg{( (a + b)*a(a + b)(a + b)(a + b)(a + b)(a + b)  $\setminus$  (a*b)*aaaaaaa*)}
\end{center}
which eliminates the symbol $\setminus$\ , giving:
\begin{center}
(a + b)*a(aaa(ab + b(a + b)) + \\(b(a + b)(a + b) + a(ba + (a + b)b))(a + b)(a + b))
\end{center}
This nice result is obtained thanks to the algorithm \verb%S% for solving equations. 

To check equivalence of two regular expressions, we can compute their symmetrical difference (noted \verb%^%). For instance, simplifying 
\begin{center}
\reg{(((xy* + yx)*} $\&$ \reg{(y*x + xy)*)} \verb%^% \reg{(yx)*(x + xy(yy*x)*)*)}
\end{center}
returns $0$. (The character $\&$ is used to represent the operation $\cap$.) 
More interesting examples can be found in the  folder \verb%testdata% of the dropbox file.

\section{More Statistics}

In this section, I present more extensive statistics on the results produced by the system on ``large'' expressions.
Four files of expressions of length 1000, using 1, 2, 3, and 4 letters have been generated in a fair way, giving to every expression the same probability to be chosen. (The files can be found in the folder \verb%testdata% of the dropbox file.)
The system was run on those files with various parametrizations. Tables \ref{sreb} to \ref{sred} gather the statistics. These tables have an additional column $\textit{gc}_f$ that counts the number of garbage collector calls that were unable to reclaim enough identifiers to complety process a sub-expression. In that case, the garbage collector is called again but the sub-expressions remaining on the stack only are pre-simplified. All executions use $1,000,000$ identifiers. Using less identifiers can be preferable on some computers.
The new tables present the same kind of statistics than Table \ref{sre} but most combinations of algorithms are considered. ``Standard'' combinations do a little more than appying the algorithms in the list to all sub-expressions: they minimize the set of all equations, just once, in the end (after complete simplification by other algorithms) and they also try a last factorization of the result. The letter \verb%n% in the first column indicates that this last attempt to simplify is not made. Other possibilities of letters have been explained before for Table \ref{sre}.

Let us have a look at Table  \ref{sreb}, first. We can see that good combinations of algorithms must include \verb%s% or \verb%rS%. Systematic factorization (\verb%f%) brings a small improvement at a relatively high cost.  The final minimization and factorisation gives an improvement mostly when nor \verb%s% nor \verb%rS% are used. The lines \verb%nra% and \verb%nraS% show what happens if a complete minimization of the set of equations is applied for every sub-expression: the execution time grows unacceptably without bringing better results. Finally, we see that most combinations minimize $25$ expressions (to \reg{(a + b)*).} Since the expressions have been chosen fairly, this suggests that $25\%$ of all expressions with two letters are equivalent to  \reg{(a + b)*.}

Table \ref{srea} presents the same statistics for expressions using only one letter. Almost all combinations of algorithms give the same (or almost the same) results. The combinations \verb%nra% and \verb%nraS% give precise results with good execution times. This is probably because the program can detect that the set of all equations was not modified since the last call to the minimization algorithm, making its current execution useless. The values in the column $n_\textit{min}$ suggest that $2/3$ of all expressions with one letter are simplifiable to a*.

Now let us consider Tables \ref{srec} and \ref{sred}. For three letters, only seven expressions are found equivalent to (a + b + c)*. For four letters, no similar result is reported. Globally, this suggests that when the number of letters increases the proportion of simplifiable expressions decreases quickly. It does not actually mean that my program is less able to simplify expressions with many letters: one can only simplify  what is simplifiable. With respect to the execution time, a similar remark seems sensible. For one or two letters, the pre-simplification of the sub-expressions often gives short expressions that can be processed quickly by the algorithms. With more letters, pre-simplified expressions are bigger, explaining the greater execution times.

\begin{table}
\caption{Simplification of random expressions of length 1000 with two letters}\label{sreb}
\begin{center}
\begin{tabular}{|r|r|r|r|r|r|r|r|r|r|r|r|r|r|r|r|r|r|}
\hline
\textit{algorithms}&& $l_N$ && $n_\textit{min}$ && $l_\textit{avg}$ & $l_{1/4}$ & $l_{1/2}$ 
& $l_{3/4}$ && $t_\textit{avg}$ & $t_{1/4}$ & $t_{1/2}$ 
& $t_{3/4}$ && \textit{gc} & $\textit{gc}_f$  \\[0.3em]\hline\hline
n && 715 && 0 && 620 & 613 & 637 & 656 && 0.44 & 0.15 & 0.33 & 0.59 && 7 & 0 \\\hline
nr && 715 && 0 && 391 & 300 & 415 & 525 && 0.2 & 0.03 & 0.08 & 0.22 && 3 & 0 \\\hline
ns && 715 && 25 && 76 & 4 & 31 & 114 && 0.11 & 0.02 & 0.03 & 0.06 && 1 & 0 \\\hline
nS && 715 && 0 && 386 & 303 & 413 & 466 && 0.27 & 0.08 & 0.17 & 0.35 && 3 & 0 \\\hline
nrs && 714 && 25 && 73 & 4 & 31 & 114 && 0.09 & 0.02 & 0.03 & 0.08 && 1 & 0 \\\hline
nrS && 714 && 25 && 54 & 4 & 20 & 64 && 0.11 & 0.04 & 0.06 & 0.1 && 1 & 0 \\\hline
nsS && 714 && 25 && 51 & 4 & 20 & 65 && 0.12 & 0.04 & 0.06 & 0.12 && 1 & 0 \\\hline
nrsS && 714 && 25 && 49 & 4 & 20 & 58 && 0.13 & 0.04 & 0.06 & 0.1 && 1 & 0 \\\hline\hline
nf && 715 && 0 && 154 & 48 & 109 & 230 && 0.95 & 0.12 & 0.23 & 0.7 && 20 & 6 \\\hline
nfr && 715 && 0 && 143 & 45 & 86 & 219 && 0.99 & 0.09 & 0.23 & 0.48 && 20 & 6 \\\hline
nfs && 715 && 25 && 58 & 4 & 25 & 80 && 0.48 & 0.05 & 0.11 & 0.3 && 10 & 0 \\\hline
nfS && 715 && 14 && 96 & 17 & 51 & 142 && 0.78 & 0.08 & 0.17 & 0.44 && 16 & 2 \\\hline
nfrs && 715 && 25 && 58 & 4 & 25 & 80 && 0.47 & 0.06 & 0.11 & 0.31 && 10 & 0 \\\hline
nfrS && 715 && 25 && 49 & 4 & 20 & 53 && 0.46 & 0.06 & 0.11 & 0.24 && 8 & 0 \\\hline
nfsS && 715 && 25 && 48 & 4 & 20 & 64 && 0.4 & 0.07 & 0.13 & 0.25 && 8 & 0 \\\hline
nfrsS && 715 && 25 && 47 & 4 & 19 & 50 && 0.42 & 0.06 & 0.12 & 0.28 && 8 & 0 \\\hline\hline
nra && 714 && 25 && 64 & 4 & 31 & 88 && 7.21 & 3.11 & 6.77 & 11.17 && 0 & 0 \\\hline
nraS && 714 && 25 && 53 & 4 & 20 & 64 && 19.4 & 6.16 & 14.05 & 28.27 && 1 & 0 \\\hline\hline
 && 715 && 24 && 269 & 5 & 59 & 614 && 0.94 & 0.28 & 0.53 & 1.08 && 6 & 0 \\\hline
r && 715 && 0 && 335 & 120 & 348 & 513 && 0.41 & 0.04 & 0.12 & 0.49 && 2 & 0 \\\hline
s && 715 && 25 && 69 & 4 & 24 & 91 && 0.18 & 0.07 & 0.1 & 0.13 && 2 & 0 \\\hline
S && 715 && 25 && 178 & 4 & 39 & 412 && 0.52 & 0.13 & 0.25 & 0.54 && 2 & 0 \\\hline
rs && 715 && 25 && 70 & 4 & 31 & 96 && 0.15 & 0.02 & 0.03 & 0.09 && 3 & 0 \\\hline
rS && 714 && 25 && 52 & 4 & 20 & 52 && 0.15 & 0.04 & 0.07 & 0.14 && 2 & 0 \\\hline
sS && 715 && 25 && 49 & 4 & 19 & 50 && 0.23 & 0.12 & 0.18 & 0.26 && 1 & 0 \\\hline
rsS && 715 && 25 && 49 & 4 & 19 & 50 && 0.19 & 0.04 & 0.07 & 0.18 && 1 & 0 \\\hline\hline
f && 715 && 25 && 102 & 4 & 47 & 136 && 1.14 & 0.15 & 0.28 & 0.72 && 19 & 5 \\\hline
fr && 715 && 6 && 127 & 36 & 81 & 202 && 1.07 & 0.12 & 0.24 & 0.6 && 19 & 5 \\\hline
fs && 715 && 25 && 58 & 4 & 20 & 73 && 0.54 & 0.11 & 0.21 & 0.37 && 10 & 0 \\\hline
fS && 715 && 25 && 67 & 4 & 31 & 91 && 0.91 & 0.16 & 0.26 & 0.5 && 17 & 1 \\\hline
frs && 715 && 25 && 59 & 4 & 24 & 73 && 0.52 & 0.07 & 0.15 & 0.34 && 10 & 0 \\\hline
frS && 715 && 25 && 49 & 4 & 20 & 52 && 0.5 & 0.09 & 0.2 & 0.4 && 9 & 0 \\\hline
fsS && 715 && 25 && 47 & 4 & 18 & 47 && 0.48 & 0.17 & 0.27 & 0.41 && 8 & 0 \\\hline
frsS && 715 && 25 && 47 & 4 & 19 & 48 && 0.42 & 0.09 & 0.17 & 0.32 && 8 & 0 \\\hline
\end{tabular}
\end{center}
\end{table}

\begin{table}
\caption{Simplification of random expressions of length 1000 with one letter}\label{srea}
\begin{center}
\begin{tabular}{|r|r|r|r|r|r|r|r|r|r|r|r|r|r|r|r|r|r|}
\hline
\textit{algorithms}&& $l_N$ && $n_\textit{min}$ && $l_\textit{avg}$ & $l_{1/4}$ & $l_{1/2}$ 
& $l_{3/4}$ && $t_\textit{avg}$ & $t_{1/4}$ & $t_{1/2}$ 
& $t_{3/4}$ && \textit{gc} & $\textit{gc}_f$  \\[0.3em]\hline\hline
n && 564 && 0 && 359 & 336 & 356 & 381 && 0.02 & 0.01 & 0.02 & 0.02 && 0 & 0 \\\hline
nr && 564 && 0 && 358 & 334 & 356 & 379 && 0.02 & 0.01 & 0.02 & 0.02 && 0 & 0 \\\hline
ns && 562 && 66 && 4 & 2 & 2 & 4 && 0.01 & 0.01 & 0.01 & 0.01 && 0 & 0 \\\hline
nS && 562 && 66 && 3 & 2 & 2 & 4 && 0.01 & 0.01 & 0.01 & 0.01 && 0 & 0 \\\hline
nrs && 562 && 66 && 4 & 2 & 2 & 4 && 0.01 & 0.01 & 0.01 & 0.01 && 0 & 0 \\\hline
nrS && 562 && 66 && 3 & 2 & 2 & 4 && 0.01 & 0.01 & 0.01 & 0.01 && 0 & 0 \\\hline
nsS && 562 && 66 && 3 & 2 & 2 & 4 && 0.01 & 0.01 & 0.01 & 0.01 && 0 & 0 \\\hline
nrsS && 562 && 66 && 3 & 2 & 2 & 4 && 0.01 & 0.01 & 0.01 & 0.01 && 0 & 0 \\\hline\hline
nf && 562 && 59 && 6 & 2 & 2 & 4 && 0.01 & 0.01 & 0.01 & 0.01 && 0 & 0 \\\hline
nfr && 562 && 59 && 6 & 2 & 2 & 4 && 0.01 & 0.01 & 0.01 & 0.01 && 0 & 0 \\\hline
nfs && 562 && 66 && 4 & 2 & 2 & 4 && 0.01 & 0.01 & 0.01 & 0.01 && 0 & 0 \\\hline
nfS && 562 && 66 && 3 & 2 & 2 & 4 && 0.01 & 0.01 & 0.01 & 0.01 && 0 & 0 \\\hline
nfrs && 562 && 66 && 4 & 2 & 2 & 4 && 0.01 & 0.01 & 0.01 & 0.01 && 0 & 0 \\\hline
nfrS && 562 && 66 && 3 & 2 & 2 & 4 && 0.01 & 0.01 & 0.01 & 0.01 && 0 & 0 \\\hline
nfsS && 562 && 66 && 3 & 2 & 2 & 4 && 0.01 & 0.01 & 0.01 & 0.01 && 0 & 0 \\\hline
nfrsS && 562 && 66 && 3 & 2 & 2 & 4 && 0.01 & 0.01 & 0.01 & 0.01 && 0 & 0 \\\hline\hline
nra && 562 && 66 && 3 & 2 & 2 & 4 && 0.01 & 0.01 & 0.01 & 0.01 && 0 & 0 \\\hline
nraS && 562 && 66 && 3 & 2 & 2 & 4 && 0.01 & 0.01 & 0.01 & 0.01 && 0 & 0 \\\hline\hline
 && 562 && 66 && 5 & 2 & 2 & 4 && 0.01 & 0.01 & 0.01 & 0.01 && 0 & 0 \\\hline
r && 563 && 20 && 91 & 4 & 20 & 187 && 0.02 & 0.01 & 0.01 & 0.02 && 0 & 0 \\\hline
s && 562 && 66 && 3 & 2 & 2 & 4 && 0.01 & 0.01 & 0.01 & 0.01 && 0 & 0 \\\hline
S && 562 && 66 && 3 & 2 & 2 & 4 && 0.01 & 0.01 & 0.01 & 0.01 && 0 & 0 \\\hline
rs && 562 && 66 && 4 & 2 & 2 & 4 && 0.01 & 0.01 & 0.01 & 0.01 && 0 & 0 \\\hline
rS && 562 && 66 && 3 & 2 & 2 & 4 && 0.01 & 0.01 & 0.01 & 0.01 && 0 & 0 \\\hline
sS && 562 && 66 && 3 & 2 & 2 & 4 && 0.01 & 0.01 & 0.01 & 0.01 && 0 & 0 \\\hline
rsS && 562 && 66 && 3 & 2 & 2 & 4 && 0.01 & 0.01 & 0.01 & 0.01 && 0 & 0 \\\hline\hline
f && 562 && 66 && 3 & 2 & 2 & 4 && 0.01 & 0.01 & 0.01 & 0.01 && 0 & 0 \\\hline
fr && 562 && 62 && 5 & 2 & 2 & 4 && 0.01 & 0.01 & 0.01 & 0.01 && 0 & 0 \\\hline
fs && 562 && 66 && 3 & 2 & 2 & 4 && 0.01 & 0.01 & 0.01 & 0.01 && 0 & 0 \\\hline
fS && 562 && 66 && 3 & 2 & 2 & 4 && 0.01 & 0.01 & 0.01 & 0.01 && 0 & 0 \\\hline
frs && 562 && 66 && 4 & 2 & 2 & 4 && 0.01 & 0.01 & 0.01 & 0.01 && 0 & 0 \\\hline
frS && 562 && 66 && 3 & 2 & 2 & 4 && 0.01 & 0.01 & 0.01 & 0.01 && 0 & 0 \\\hline
fsS && 562 && 66 && 3 & 2 & 2 & 4 && 0.01 & 0.01 & 0.01 & 0.01 && 0 & 0 \\\hline
frsS && 562 && 66 && 3 & 2 & 2 & 4 && 0.01 & 0.01 & 0.01 & 0.01 && 0 & 0 \\\hline

\end{tabular}
\end{center}
\end{table}

\begin{table}
\caption{Simplification of random expressions of length 1000 with three letters}\label{srec}
\begin{center}
\begin{tabular}{|r|r|r|r|r|r|r|r|r|r|r|r|r|r|r|r|r|r|}
\hline
\textit{algorithms}&& $l_N$ && $n_\textit{min}$ && $l_\textit{avg}$ & $l_{1/4}$ & $l_{1/2}$ 
& $l_{3/4}$ && $t_\textit{avg}$ & $t_{1/4}$ & $t_{1/2}$ 
& $t_{3/4}$ && \textit{gc} & $\textit{gc}_f$  \\[0.3em]\hline\hline
n && 783 && 0 && 729 & 718 & 733 & 751 && 1.51 & 0.26 & 0.53 & 1.49 && 19 & 0 \\\hline
nr && 783 && 0 && 596 & 517 & 678 & 721 && 1.07 & 0.12 & 0.38 & 0.75 && 13 & 0 \\\hline
ns && 783 && 7 && 315 & 97 & 321 & 516 && 2.89 & 0.13 & 0.27 & 1.16 && 21 & 3 \\\hline
nS && 783 && 0 && 620 & 566 & 632 & 668 && 1.44 & 0.28 & 0.47 & 1.36 && 17 & 0 \\\hline
nrs && 783 && 7 && 315 & 97 & 321 & 513 && 2.92 & 0.14 & 0.29 & 1.16 && 21 & 3 \\\hline
nrS && 783 && 7 && 311 & 79 & 283 & 518 && 0.79 & 0.11 & 0.23 & 0.54 && 8 & 0 \\\hline
nsS && 783 && 7 && 285 & 79 & 267 & 467 && 2.64 & 0.21 & 0.46 & 1.11 && 18 & 1 \\\hline
nrsS && 783 && 7 && 281 & 78 & 267 & 467 && 2.61 & 0.19 & 0.46 & 1.11 && 18 & 1 \\\hline\hline
nf && 783 && 0 && 483 & 356 & 552 & 627 && 5.84 & 1.05 & 3.5 & 7.85 && 73 & 30 \\\hline
nfr && 783 && 1 && 461 & 309 & 547 & 626 && 5.77 & 0.99 & 3.39 & 7.96 && 72 & 29 \\\hline
nfs && 783 && 7 && 305 & 93 & 314 & 492 && 3.57 & 0.64 & 2.11 & 5.19 && 46 & 18 \\\hline
nfS && 783 && 0 && 358 & 189 & 371 & 527 && 4.2 & 0.8 & 2.58 & 6.09 && 54 & 18 \\\hline
nfrs && 783 && 7 && 305 & 93 & 314 & 492 && 3.63 & 0.66 & 2.2 & 5.13 && 46 & 18 \\\hline
nfrS && 783 && 7 && 293 & 89 & 279 & 480 && 3.27 & 0.69 & 1.68 & 5.21 && 39 & 17 \\\hline
nfsS && 783 && 7 && 282 & 79 & 266 & 449 && 3.37 & 0.69 & 1.86 & 5.15 && 42 & 16 \\\hline
nfrsS && 783 && 7 && 280 & 79 & 266 & 446 && 3.37 & 0.68 & 2.03 & 4.86 && 41 & 16 \\\hline\hline
nra && 783 && 5 && 373 & 110 & 394 & 583 && 33.86 & 9.0 & 24.99 & 46.37 && 8 & 0 \\\hline
nraS && 783 && 7 && 310 & 79 & 283 & 516 && 69.77 & 19.26 & 59.45 & 102.78 && 8 & 0 \\\hline\hline
 && 783 && 3 && 643 & 694 & 722 & 741 && 3.43 & 0.77 & 1.49 & 3.15 && 12 & 0 \\\hline
r && 783 && 0 && 591 & 516 & 670 & 722 && 2.18 & 0.42 & 0.88 & 1.83 && 10 & 0 \\\hline
s && 783 && 7 && 313 & 96 & 319 & 504 && 4.0 & 0.41 & 0.67 & 1.71 && 18 & 3 \\\hline
S && 783 && 7 && 554 & 550 & 622 & 666 && 2.91 & 0.79 & 1.38 & 3.0 && 11 & 0 \\\hline
rs && 783 && 7 && 313 & 96 & 319 & 511 && 3.86 & 0.22 & 0.58 & 1.67 && 18 & 3 \\\hline
rS && 783 && 7 && 307 & 78 & 281 & 523 && 1.62 & 0.23 & 0.59 & 1.51 && 7 & 0 \\\hline
sS && 783 && 7 && 283 & 79 & 260 & 466 && 3.37 & 0.5 & 0.9 & 1.9 && 18 & 1 \\\hline
rsS && 783 && 7 && 280 & 78 & 260 & 466 && 3.27 & 0.36 & 0.71 & 1.87 && 17 & 1 \\\hline\hline
f && 783 && 4 && 442 & 263 & 544 & 624 && 7.48 & 1.5 & 4.02 & 9.48 && 72 & 30 \\\hline
fr && 783 && 1 && 454 & 281 & 547 & 620 && 7.61 & 1.52 & 3.66 & 9.31 && 71 & 29 \\\hline
fs && 783 && 7 && 305 & 93 & 311 & 487 && 4.24 & 0.79 & 2.38 & 6.11 && 45 & 18 \\\hline
fS && 783 && 7 && 333 & 172 & 350 & 527 && 5.01 & 1.05 & 2.7 & 6.19 && 54 & 18 \\\hline
frs && 783 && 7 && 305 & 93 & 314 & 489 && 4.15 & 0.79 & 2.19 & 5.94 && 45 & 18 \\\hline
frS && 783 && 7 && 292 & 89 & 279 & 478 && 4.0 & 0.91 & 2.09 & 5.74 && 38 & 17 \\\hline
fsS && 783 && 7 && 282 & 79 & 266 & 445 && 4.15 & 0.89 & 2.13 & 5.29 && 42 & 16 \\\hline
frsS && 783 && 7 && 279 & 79 & 266 & 442 && 4.08 & 0.9 & 2.22 & 5.18 && 41 & 16 \\\hline
\end{tabular}
\end{center}
\end{table}

\begin{table}
\caption{Simplification of random expressions of length 1000 with four letters}\label{sred}
\begin{center}
\begin{tabular}{|r|r|r|r|r|r|r|r|r|r|r|r|r|r|r|r|r|r|}
\hline
\textit{algorithms}&& $l_N$ && $n_\textit{min}$ && $l_\textit{avg}$ & $l_{1/4}$ & $l_{1/2}$ 
& $l_{3/4}$ && $t_\textit{avg}$ & $t_{1/4}$ & $t_{1/2}$ 
& $t_{3/4}$ && \textit{gc} & $\textit{gc}_f$  \\[0.3em]\hline\hline
n && 825 && 0 && 791 & 782 & 796 & 811 && 0.56 & 0.07 & 0.18 & 0.48 && 7 & 0 \\\hline
nr && 825 && 0 && 743 & 758 & 784 & 802 && 0.56 & 0.1 & 0.25 & 0.46 && 7 & 0 \\\hline
ns && 825 && 0 && 544 & 434 & 650 & 700 && 1.76 & 0.22 & 0.54 & 1.69 && 14 & 1 \\\hline
nS && 825 && 0 && 731 & 715 & 734 & 761 && 2.15 & 0.28 & 0.4 & 0.76 && 8 & 0 \\\hline
nrs && 825 && 0 && 544 & 434 & 650 & 700 && 1.81 & 0.25 & 0.58 & 1.74 && 14 & 1 \\\hline
nrS && 825 && 0 && 564 & 444 & 681 & 734 && 1.74 & 0.22 & 0.41 & 0.7 && 7 & 0 \\\hline
nsS && 825 && 0 && 531 & 416 & 645 & 688 && 2.98 & 0.42 & 0.88 & 1.91 && 15 & 1 \\\hline
nrsS && 825 && 0 && 528 & 416 & 640 & 688 && 3.29 & 0.44 & 0.91 & 1.94 && 15 & 1 \\\hline\hline
nf && 825 && 0 && 660 & 668 & 709 & 733 && 9.02 & 2.63 & 7.07 & 11.61 && 71 & 26 \\\hline
nfr && 825 && 0 && 663 & 673 & 710 & 732 && 9.05 & 2.68 & 7.17 & 11.62 && 72 & 26 \\\hline
nfs && 825 && 0 && 526 & 420 & 621 & 681 && 8.25 & 2.04 & 4.83 & 11.57 && 61 & 17 \\\hline
nfS && 825 && 0 && 591 & 568 & 661 & 700 && 9.54 & 2.74 & 7.28 & 11.02 && 66 & 22 \\\hline
nfrs && 825 && 0 && 526 & 420 & 621 & 681 && 7.73 & 1.98 & 4.56 & 10.48 && 60 & 17 \\\hline
nfrS && 825 && 0 && 538 & 475 & 639 & 695 && 7.98 & 2.2 & 4.95 & 10.56 && 58 & 19 \\\hline
nfsS && 825 && 0 && 520 & 418 & 616 & 673 && 8.78 & 2.22 & 5.03 & 11.16 && 60 & 18 \\\hline
nfrsS && 825 && 0 && 524 & 451 & 616 & 673 && 8.21 & 2.17 & 4.94 & 10.69 && 60 & 19 \\\hline\hline
nra && 825 && 0 && 624 & 535 & 744 & 777 && 199.37 & 52.23 & 191.99 & 308.54 && 5 & 0 \\\hline
nraS && 825 && 0 && 562 & 440 & 681 & 732 && 338.69 & 56.16 & 259.75 & 518.1 && 7 & 0 \\\hline\hline
 && 825 && 0 && 766 & 775 & 790 & 809 && 2.52 & 0.78 & 1.5 & 2.98 && 4 & 0 \\\hline
r && 825 && 0 && 742 & 759 & 782 & 802 && 2.05 & 0.44 & 1.16 & 2.56 && 4 & 0 \\\hline
s && 825 && 0 && 541 & 434 & 650 & 700 && 3.27 & 0.99 & 1.78 & 3.7 && 9 & 1 \\\hline
S && 825 && 0 && 704 & 709 & 734 & 756 && 3.61 & 1.31 & 2.28 & 3.68 && 1 & 0 \\\hline
rs && 825 && 0 && 541 & 434 & 650 & 700 && 2.61 & 0.59 & 1.23 & 3.32 && 9 & 1 \\\hline
rS && 825 && 0 && 563 & 444 & 684 & 734 && 3.3 & 0.78 & 1.57 & 3.47 && 4 & 0 \\\hline
sS && 825 && 0 && 528 & 416 & 645 & 688 && 4.67 & 1.53 & 2.65 & 4.42 && 11 & 1 \\\hline
rsS && 825 && 0 && 527 & 416 & 636 & 688 && 4.67 & 1.16 & 1.87 & 3.77 && 10 & 1 \\\hline\hline
f && 825 && 0 && 658 & 668 & 707 & 732 && 10.87 & 3.72 & 7.59 & 13.25 && 70 & 26 \\\hline
fr && 825 && 0 && 661 & 670 & 706 & 731 && 10.84 & 3.54 & 8.06 & 14.12 && 71 & 26 \\\hline
fs && 825 && 0 && 530 & 456 & 621 & 681 && 9.19 & 2.83 & 5.38 & 11.81 && 61 & 18 \\\hline
fS && 825 && 0 && 584 & 549 & 661 & 700 && 10.47 & 3.85 & 7.36 & 13.31 && 66 & 22 \\\hline
frs && 825 && 0 && 526 & 420 & 621 & 681 && 9.29 & 2.68 & 5.4 & 12.65 && 60 & 17 \\\hline
frS && 825 && 0 && 538 & 475 & 639 & 695 && 9.86 & 3.4 & 6.05 & 11.7 && 58 & 19 \\\hline
fsS && 825 && 0 && 524 & 451 & 616 & 673 && 10.06 & 3.32 & 5.9 & 11.55 && 60 & 19 \\\hline
frsS && 825 && 0 && 520 & 418 & 616 & 673 && 9.9 & 3.31 & 5.75 & 11.73 && 59 & 18 \\\hline
\end{tabular}
\end{center}
\end{table}

\subsubsection{Acknowledgements} I want to thank Yves Deville, Pierre Flener, and Jos\'e Vander Meulen for their interest in my work and their useful comments.

\end{document}